\documentclass[sigconf]{acmart}
\usepackage{hyperref}
\usepackage{breakurl}
\AtBeginDocument{%
  \providecommand\BibTeX{{%
    \normalfont B\kern-0.5em{\scshape i\kern-0.25em b}\kern-0.8em\TeX}}}

\copyrightyear{2023}
\acmYear{2023}
\setcopyright{acmlicensed}
\acmConference[W4A '23]{20th International Web for All Conference}{April 30-May 1, 2023}{Austin, TX, USA}
\acmBooktitle{20th International Web for All Conference (W4A '23), April 30-May 1, 2023, Austin, TX, USA}
\acmPrice{15.00}
\acmDOI{10.1145/3587281.3587282}
\acmISBN{979-8-4007-0748-3/23/04}

\acmSubmissionID{4}
\begin{document}

\title{Accessibility Metatesting}
\subtitle{Comparing Nine Testing Tools}

\author{Jonathan Robert Pool}
\email{jonathan.pool@cvshealth.com}
\orcid{0000-0001-7864-5229}
\affiliation{%
  \institution{CVS Health}
  \streetaddress{1 CVS Drive}
  \city{Woonsocket}
  \state{Rhode Island}
  \country{USA}
  \postcode{02895}
}
\begin{abstract}
Automated web accessibility testing tools have been found complementary. The implication: To catch as many issues as possible, use multiple tools. Doing this efficiently entails integration costs. Is there a small set of tools that, together, make additional tools redundant? I approach this problem by comparing nine comprehensive accessibility testing tools that are amenable to integration: alfa, axe-core, Continuum, Equal Access, HTML CodeSniffer, Nu Html Checker, QualWeb, Tenon, and WAVE. I tested 121 web pages of interest to CVS Health with these tools. Each tool only fractionally duplicated any other tool. Each discovered numerous issue instances missed by all the others. Thus, testing with all nine tools was substantially more informative than testing with any subset.
\end{abstract}

\begin{CCSXML}
<ccs2012>
   <concept>
       <concept_id>10011007.10011074.10011099.10011693</concept_id>
       <concept_desc>Software and its engineering~Empirical software validation</concept_desc>
       <concept_significance>300</concept_significance>
       </concept>
   <concept>
       <concept_id>10011007.10011074.10011099.10011102.10011103</concept_id>
       <concept_desc>Software and its engineering~Software testing and debugging</concept_desc>
       <concept_significance>500</concept_significance>
       </concept>
   <concept>
       <concept_id>10003120.10011738.10011773</concept_id>
       <concept_desc>Human-centered computing~Empirical studies in accessibility</concept_desc>
       <concept_significance>300</concept_significance>
       </concept>
   <concept>
       <concept_id>10003120.10011738.10011774</concept_id>
       <concept_desc>Human-centered computing~Accessibility design and evaluation methods</concept_desc>
       <concept_significance>500</concept_significance>
       </concept>
   <concept>
       <concept_id>10003120.10011738.10011775</concept_id>
       <concept_desc>Human-centered computing~Accessibility technologies</concept_desc>
       <concept_significance>500</concept_significance>
       </concept>
   <concept>
       <concept_id>10003120.10011738.10011776</concept_id>
       <concept_desc>Human-centered computing~Accessibility systems and tools</concept_desc>
       <concept_significance>500</concept_significance>
       </concept>
 </ccs2012>
\end{CCSXML}

\ccsdesc[300]{Software and its engineering~Empirical software validation}
\ccsdesc[500]{Software and its engineering~Software testing and debugging}
\ccsdesc[300]{Human-centered computing~Empirical studies in accessibility}
\ccsdesc[500]{Human-centered computing~Accessibility design and evaluation methods}
\ccsdesc[500]{Human-centered computing~Accessibility technologies}
\ccsdesc[500]{Human-centered computing~Accessibility systems and tools}
\keywords{web accessibility, accessibility testing, metatesting, test automation, test efficiency}

\received{17 January 2023}
\received[accepted]{28 February 2023}
\received[revised]{14 March 2023}
\received[revised]{12 April 2023}
\maketitle

\section{Introduction}

Organizations that test their own, their competitors', their suppliers', and other web assets for accessibility \cite{W3C1} can handle part of the testing workload with automated and semi-automated tools. The World Wide Web Consortium (W3C) lists 167 such tools \cite{W3C2}. Some are designed for the fully automated and comprehensive discovery of accessibility issues. They include web services, APIs, browser extensions, and installable software.

Comparisons of such tools have found them substantially complementary: Issues discovered by one are often overlooked by another. This has led to recommendations to use multiple tools rather than relying on only one \cite{Abdu}\cite{Pad}.

The more tools one uses, the greater the cost of integrating them into a regime of automated accessibility testing. This motivates the question, how many, and which, tools to deploy. The most obvious candidates are free or nearly free tools that conform to standard protocols, can be controlled programmatically, and aim to test accessibility comprehensively.

I explore the tool-selection problem here by studying the issues reported by nine accessibility testing tools amenable to integration.

\section{Related Work}

Abduganiev \cite{Abdu} compared eight accessibility testing tools on fifty-two web pages from Tajikistan and Austria, and Pădure and Pribeanu \cite{Pad} compared six accessibility testing tools on the websites of six municipal governments in Romania. Both reported enough complementarity to justify using multiple tools.

Ara and Sik-Lanyi \cite{Ara} used four tools to test the accessibility of twenty COVID-19 vaccination websites of governments of mostly European nation states. They found large inter-tool differences in reported error counts. For example, the Mauve tool found 12 errors on the Dutch site and 16 errors on the German site, but the Web Accessibility tool found 104 errors on the Dutch site and 1 error on the German site.

Burkard, Zimmermann, and Schwarzer \cite{Burk} evaluated four commercial accessibility-monitoring tools. The one that discovered the most issues did not render redundant the one that discovered the fewest. The latter still discovered violations of six accessibility standards that the former did not discover \cite{BurkSupp}.

Silva, Oliveira, Mateus, Costa, and Freire \cite{Silva} reviewed four studies and tabulated the accessibility issues reported by the tools they had employed. The analysis enables a comparison of only two tools: SortSite and WAVE. One issue was reported by both; one was reported by only SortSite; and nine were reported by only WAVE. Thus, the tools were almost totally complementary.

Kumar, Biswas, and Venkatesh \cite{Kumar} tested the websites of the World Health Organization and the British Broadcasting Corporation with ten tools and reported that five of them (A-Checker, Functional Accessibility Evaluator, Accessibility Insights, Utilitia Validator, and IBM Accessibility Equal Access Toolkit) covered all four Principles of the Web Content Accessibility Guidelines. These five tools reported, among them, 11 issues. Of these, 8 were reported by 1 tool, 2 by 2 tools, and 1 by 3 tools. To capture all 11 issues, it would be necessary to use at least 3 of the tools.

The reviewed research supports the generalization that purportedly comprehensive accessibility testing tools differ substantially in the issues they report.

\section{Tools}

I classify tools as amenable to integration into a multi-tool regime of automated testing if they are currently maintained, free or nearly free, and either usable as REST APIs or installable as NPM packages. By this classification, of the fourteen tools used in the studies cited above, two, and derivatives of two more, are amenable to integration. I have chosen these four and five additional qualifying tools for comparison here. The tools I compare are listed in Table~\ref{tab:pkgs}. They all seek to measure conformance to industry standards \cite{W3C3} and best practices for accessibility, so it is reasonable to speculate that some of them might make some others of them redundant.

The nine tools compared here should not be expected to remain viable candidates, or the only viable candidates, because, as Abduganiev \cite{Abdu} notes, tools rapidly appear and disappear. In fact, one of the tools I compare (Tenon) became unavailable for new subscribers after I conducted this research.

\begin{table}
  \caption{Accessibility Testing Tools}
  \label{tab:pkgs}
  \begin{tabular}{lllr}
    \toprule
    Code & Name & Creator & Tests\\
    \midrule
    alfa & alfa \cite{alfa}& Siteimprove & 103\\
    axe & axe-core \cite{axe}& Deque & 138\\
    continuum & Continuum \cite{continuum}& Level Access & 267\\
    htmlcs & HTML CodeSniffer \cite{htmlcs}& Squiz & 98\\
    ibm & Equal Access \cite{ibm}& IBM & 163\\
    nuVal & Nu Html Checker \cite{nuVal}& W3C & 147\\
    qualWeb & QualWeb \cite{qualWeb}& Universidade da Lisboa & 121\\
    tenon & Tenon \cite{tenon}& Tenon.io & 180\\
    wave & WAVE \cite{wave}& WebAIM & 110\\
    \bottomrule
    Total & & & 1,327
  \end{tabular}
\end{table}

\section{Methods}

I classified the tools' 1,327 tests into 245 ``issues'' (defects and suspected defects), such that any reports of the same issue are arguably duplicative. For example, the ``labelClash'' issue is a form control improperly having multiple labels. Defects reported as definitive are classified as different issues from mere suspicions or warnings. The supplementary materials include the complete issue classification.

One could analyze tool complementarity \textit{a priori} on the basis of such a classification to determine which tools can discover which issues. For example, 3 of the 9 tools (axe-core, IBM Equal Access, and WAVE) have tests classified as ``labelClash''.

However, an \textit{a priori} analysis would lead to decisions based on \emph{purported} tool capabilities, which might only later be found inaccurate. Instead, I approach the problem empirically by analyzing the issues that the tools actually report.

Using the case of CVS Health, I assembled a judgmental sample \cite{Westfall} of 140 web pages that an enterprise might monitor for accessibility---internal home-built, internal vendor-produced, external own, external supplier, and external competitor pages. I then removed from the sample any publicly unreachable pages and any pages requiring pre-test interaction (such as logging in), to allow externally hosted tools without scripted actions to test the whole sample. This left a sample of 121 pages.

I tested all 121 pages with all nine tools. This permitted an empirical tabulation of duplication and complementarity among the tools. With this empirical approach, a tool's presumed value is based on the issue instances that it reports but other tools do not.

This method, although empirical, is still naïve:

\begin{itemize}
\item It trusts the claims of each tool as to the issue instances it has discovered.
\item It assumes that, if on some page tool $A$ reports $m$ instances of issue $C$ and tool $B$ reports $n$ instances of issue $C$, where $n>m$, then the instances reported by $A$ are a subset of the instances reported by $B$.
\end{itemize}

Thus, in the analysis below, references to issues discovered are, strictly speaking, references to \emph{claims} of issues discovered.

My trust in the claims of tools contrasts with prior studies \cite{Abdu}\cite{Burk} that employed human testers to classify automatically reported issues as correct (``true positives'') or incorrect (``false positives'').

There are reasons for such trust. False positivity is, arguably, a judgment, not a fact. Moreover, litigators reportedly use automated tests to identify inaccessible websites \cite{Riv}\cite{Ros}; thus, treating automated failure reports as presumptively correct, thereby avoiding not only inaccessibility but the \emph{appearance} of inaccessibility, can mitigate the risk of claims and disputes.

\section{Results}

\subsection{Totals}

The tools differed substantially in how many issue instances they reported, with the most prolific tool reporting about eight times as many instances as the least prolific tool, as seen in Table~\ref{tab:totals}.

\begin{table}
  \caption{Counts of Issue Instances Reported}
  \label{tab:totals}
  \begin{tabular}{lr}
    \toprule
    Tool & Count\\
    \midrule
    qualWeb & 23,715\\
    axe & 12,364\\
    tenon & 8,328\\
    nuVal & 6,986\\
    htmlcs & 6,329\\
    ibm & 6,242\\
    wave & 6,095\\
    alfa & 5,605\\
    continuum & 3,089\\
    \bottomrule
    Total & 78,753
  \end{tabular}
\end{table}

This result could tempt one to conclude that the QualWeb tool does everything the other tools do and more, making it possible to use that tool alone without sacrificing any information, but that conclusion would not survive further analysis, as shown below.

\subsection{Pairs}

If one tool made the others redundant, then, if paired with each other tool, it would, on every tested page, report at least as many instances of each issue as the other tool reports.

In fact, however, with each of the 72 pairs of tools, each tool found, on one or more pages, more instances of some issues than the other tool found. For example, the alfa tool found 4,652 instances that the Continuum tool missed, but Continuum found 2,136 instances that alfa missed. Thus, for any two tools, testing with both revealed more instances than testing with either tool alone.

The pairwise percentage increases of issue instances from testing with two tools instead of one are shown in Table~\ref{tab:instanceMore}. For example, the ``48'' in the bottom row means that testing my sample with both WAVE and Continuum Community Edition produced reports of 48\% more issue instances than testing with only WAVE. Adding a second tool increased the count by anywhere from 13\% to 767\%.

\begin{table}
  \caption{Pairwise Increase in Issue Instance Count (\%)}
  \label{tab:instanceMore}
  \begin{tabular}{lrrrrrrrrr}
    \toprule
    & & & & 2nd\\
    1st & alf & axe & con & htm & ibm & nuV & qW & ten & wav\\
    \midrule
    alfa & - & 193 & 38 & 105 & 96 & 123 & 402 & 143 & 100\\
    axe & 33 & - & 16 & 43 & 38 & 54 & 188 & 63 & 39\\
    continuum & 151 & 366 & - & 169 & 169 & 218 & 767 & 265 & 191\\
    htmlcs & 81 & 179 & 31 & - & 86 & 107 & 370 & 128 & 76\\
    ibm & 76 & 174 & 33 & 88 & - & 104 & 379 & 130 & 88\\
    nuVal & 79 & 173 & 40 & 87 & 82 & - & 339 & 119 & 85\\
    qualWeb & 19 & 50 & 13 & 25 & 26 & 29 & - & 33 & 24\\
    tenon & 64 & 142 & 35 & 74 & 73 & 84 & 279 & - & 61\\
    wave & 84 & 183 & 48 & 83 & 92 & 112 & 384 & 120 & -\\
    \bottomrule
  \end{tabular}
\end{table}

One might care how many additional issues a tool discovers, regardless of the instance count. That information, for this sample, is reported in Table~\ref{tab:issueMore}. Thus, adding a second tool added instances of anywhere from 15 issues (if Tenon was added to WAVE) to 51 issues (if Nu Html Checker was added to QualWeb).

\begin{table}
  \caption{Pairwise Increase in Issue Discovery}
  \label{tab:issueMore}
  \begin{tabular}{lrrrrrrrrr}
    \toprule
    & & & & 2nd\\
    1st & alf & axe & con & htm & ibm & nuV & qW & ten & wav\\
    \midrule
    alfa & - & 47 & 34 & 41 & 49 & 46 & 22 & 20 & 42\\
    axe & 25 & - & 25 & 40 & 43 & 40 & 21 & 20 & 42\\
    continuum & 34 & 46 & - & 41 & 46 & 42 & 25 & 21 & 44\\
    htmlcs & 31 & 39 & 28 & - & 45 & 45 & 21 & 19 & 39\\
    ibm & 29 & 40 & 26 & 36 & - & 43 & 23 & 20 & 40\\
    nuVal & 34 & 48 & 31 & 36 & 46 & - & 26 & 21 & 43\\
    qualWeb & 36 & 48 & 35 & 41 & 49 & 51 & - & 19 & 39\\
    tenon & 36 & 50 & 34 & 42 & 50 & 50 & 24 & - & 42\\
    wave & 33 & 41 & 28 & 31 & 41 & 46 & 22 & 15 & -\\
    \bottomrule
  \end{tabular}
\end{table}

\subsection{Issues}

Testing tools, to some extent, specialize in issues. If an organization cares about particular accessibility issues, it might be practical to test mainly or only with tools proficient in those issues.

The most evident specializations included:

\begin{itemize}
\item alfa: font and line sizing; skip-to-content links
\item axe-core: color contrast; landmark placement
\item Continuum: landmark purpose; placeholder versus label
\item HTML CodeSniffer: heading levels; semantic use of elements
\item Equal Access: landmark coverage; landmark purpose
\item Nu Html Checker: control placement; attribute validity
\item QualWeb: video alternatives; focus indication
\item Tenon: horizontal scrolling; link purpose
\item WAVE: label clarity; link purpose
\end{itemize}

If, however, an organization cares about all accessibility issues, selecting a subset of tools is problematic. In my sample, every tool had at least 10 issues that it found more instances of than did any other tool. For example, five of the tools found instances of a bad iframe title, but HTML CodeSniffer found 108 such instances, while no other tool found more than 27. The supplementary materials include a complete list of these specializations.

Moreover, every tool had at least 7 issues that \emph{only} that tool found instances of, as shown in Table~\ref{tab:only}. For example, only WAVE found instances of pages entirely missing landmarks, and only axe-core found instances of invisible form-control labels. The supplementary materials include a complete list of these sole-source issues.

\begin{table}
  \caption{Counts of issues Reported by Only One Tool}
  \label{tab:only}
  \begin{tabular}{lr}
    \toprule
    Tool & Count\\
    \midrule
    nuVal & 26\\
    htmlcs & 15\\
    ibm & 15\\
    axe & 12\\
    wave & 12\\
    continuum & 9\\
    alfa & 8\\
    tenon & 8\\
    qualWeb & 7\\
    \bottomrule
    Total & 112
  \end{tabular}
\end{table}

In some cases, the failure of a tool to discover instances of an issue on a page was due to the inability of that tool to test that page at all. Of 1,089 attempts to test a page with a tool, 49 (4\%) failed. A tool that successfully tests a page and does not find some issue has the same practical effect as a tool that fails to test the page: That tool does not discover that issue, but another tool might.

\subsection{Duplication}

While tools complemented each other, they also duplicated each other. The number of tools discovering issue instances ranged from 1 to 8. For example, 8 tools discovered links without accessible names. Even so, the duplication was imperfect. For example, WAVE found 240 such links, while the Nu Html Checker found only 2.

\section{Conclusion}

Although all the tools except Nu Html Checker claim to be comprehensive web-accessibility testers, with my sample of web pages each tool only fractionally duplicated any other tool. Each discovered numerous instances missed by the others, and each was the only one to discover instances of some issues. For this sample, testing with all nine tools was substantially more informative than testing with any smaller set or with only a single tool.

No single accessibility testing tool can be trusted to discover all automatically detectable defects. There is reason to expect more discovery from testing with multiple tools.

Routinely and efficiently testing for accessibility with as many as nine tools, even if free or inexpensive, involves effort to:

\begin{itemize}
\item integrate tool invocation methods
\item ensure test isolation
\item integrate reporting formats of tools
\item integrate tool severity and certainty classifications
\item integrate tool methods for identifying instance locations
\item integrate tool methods for pre-test browser actions
\item reconcile granularity differences in issue classifications
\item determine which issues should be considered identical
\item assign levels of trust to tools and their tests \cite{Y}
\item compensate for tool duplicativity when computing scores
\item gather findings on each issue from tool reports
\item keep track of the revisions of each tool
\item handle tool disagreements
\item integrate tools with custom tests and standards
\end{itemize}

In judging whether the benefits justify this effort, one can consider that the investment is, in part, nonrepeating, amortizable over multiple uses, and sharable by multiple organizations. Moreover, testing with an ensemble of tools can influence tool quality, by facilitating direct comparisons among tool behaviors, helping tool users discover and report tool deficiencies, and motivating the development of new, nonduplicative accessibility tests.

\appendix

\section{Disclaimer}

My opinions expressed herein are my own views and do not necessarily reflect the views of CVS Health, its affiliates, or any of my colleagues at CVS Health or its affiliates.

\section{Appendix: Supplementary Documents}

\subsection{Issue Classification}

The ``issueClassification.json'' file defines the issue classification used in this investigation. For each issue, the most applicable Web Content Accessibility Guidelines 2.1 \cite{W3C3} Success Criterion, Guideline, or Principle and the tests detecting the issue are given. For each test, the tool's own test ID, whether that ID is ``variable'' (i.e. is a regular expression to be matched rather than a string), and a paraphrase of the tool's description of the issue are given.

\subsection{Sole-Source Issues}

The ``onlies.json'' file documents the issues discovered by only one tool each. The count of issue instances found, an ID of the issue in the issue classification, and a paraphrase of the issue are given.

\subsection{Tool Issue Tabulation}

The ``issues.json'' file documents, in its ``issues'' object, the number of instances of each issue reported by each tool. In its ``most'' object, it also documents the issues that each tool reported more instances of than any other tool. For example, the fact that ``legendMissing'' is in the ``most.ibm.issues'' array means that the ``ibm'' tool reported more instances of the ``legendMissing'' issue than any other tool.

\section{Version}

© Jonathan Robert Pool 2023. This is a minor revision of the work published in \textit{20th International Web for All Conference (W4A ’23), April 20–May 1, 2023, Austin, TX, USA}.

\noindent
http://doi.org/10.1145/3587281.3587282

\bibliographystyle{ACM-Reference-Format}
\bibliography{accessibility-metatesting-20230412}

\end{document}